
\newif\ifproofmode			
\proofmodefalse				

\newif\ifforwardreference		
\forwardreferencetrue			

\newif\ifeqchapternumbers		
\eqchapternumbersfalse			

\newif\ifsectionnumbers			
\sectionnumberstrue			

\newif\ifeqsectionnumbers		
\eqsectionnumbersfalse			

\newif\ifchaptersectionnumbers     	
\chaptersectionnumberstrue		

\newif\ifcontinuoussectionnumbers	
\continuoussectionnumbersfalse	

\newif\ifcontinuousnumbers		
\continuousnumbersfalse 		

\newif\iffigurechapternumbers		
\figurechapternumbersfalse		

\newif\ifcontinuousfigurenumbers	
\continuousfigurenumbersfalse		

\newif\ifcontinuousreferencenumbers     
\continuousreferencenumberstrue         

\newif\ifparenequations			
\parenequationstrue			

\newif\ifstillreading			

\font\eqsixrm=cmr6			
\def\marginstyle{\eqsixrm}		

\newtoks\chapletter			
\newcount\chapno			
\newcount\sectno			
\newcount\eqlabelno			
\newcount\figureno			
\newcount\referenceno			
\newcount\minutes			
\newcount\hours				

\newread\labelfile			
\newwrite\labelfileout			
\newwrite\allcrossfile			

\chapno=0
\sectno=0
\eqlabelno=0
\figureno=0


%
\def\initialeqmacro{
    \ifproofmode
        \headline{\tenrm \today\ --\ \timeofday\hfill
                         \jobname\ --- draft\hfill\folio}
        \hoffset=-1cm
        \immediate\openout\allcrossfile=zallcrossreferfile
    \fi
    \ifforwardreference
        \openin\labelfile=zlabelfile
        \ifeof\labelfile
        \else
            \stillreadingtrue
            \loop
                \read\labelfile to \nextline
                \ifeof\labelfile
                    \stillreadingfalse
                \else
                    \nextline
                \fi
                \ifstillreading
            \repeat
        \fi
        \immediate\openout\labelfileout=zlabelfile
    \fi}


{\catcode`\^^I=9
\catcode`\ =9
\catcode`\^^M=9
\endlinechar=-1
\globaldefs=1


%
\def\chapfolio{			
    \ifnum \chapno>0 \relax
        \the\chapno
    \else
        \the\chapletter
    \fi}

%
\def\bumpchapno{
    \ifnum \chapno>-1 \relax
        \global \advance \chapno by 1
    \else
        \global \advance \chapno by -1 \setletter\chapno
    \fi
    \ifcontinuousnumbers
    \else
        \global\eqlabelno=0
    \fi
    \ifcontinuousfigurenumbers
    \else
        \global\figureno=0
    \fi
    \ifcontinuousreferencenumbers
    \else
        \global\referenceno=0
    \fi
    \sectno=0}

\def\bumpsectno{
    \global\advance\sectno by 1 \relax
    \ifeqsectionnumbers
        \ifcontinuoussectionnumbers
        \else
            \global\eqlabelno=0
        \fi
    \fi}

%
\def\setletter#1{\ifcase-#1 {}  \or\global\chapletter={A}
  \or\global\chapletter={B} \or\global\chapletter={C} \or\global\chapletter={D}
  \or\global\chapletter={E} \or\global\chapletter={F} \or\global\chapletter={G}
  \or\global\chapletter={H} \or\global\chapletter={I} \or\global\chapletter={J}
  \or\global\chapletter={K} \or\global\chapletter={L} \or\global\chapletter={M}
  \or\global\chapletter={N} \or\global\chapletter={O} \or\global\chapletter={P}
  \or\global\chapletter={Q} \or\global\chapletter={R} \or\global\chapletter={S}
  \or\global\chapletter={T} \or\global\chapletter={U} \or\global\chapletter={V}
  \or\global\chapletter={W} \or\global\chapletter={X} \or\global\chapletter={Y}
  \or\global\chapletter={Z}\fi}

%
\def\tempsetletter#1{\ifcase-#1 {}\or{} \or\chapletter={A} \or\chapletter={B}
 \or\chapletter={C} \or\chapletter={D} \or\chapletter={E}
  \or\chapletter={F} \or\chapletter={G} \or\chapletter={H}
   \or\chapletter={I} \or\chapletter={J} \or\chapletter={K}
    \or\chapletter={L} \or\chapletter={M} \or\chapletter={N}
     \or\chapletter={O} \or\chapletter={P} \or\chapletter={Q}
      \or\chapletter={R} \or\chapletter={S} \or\chapletter={T}
       \or\chapletter={U} \or\chapletter={V} \or\chapletter={W}
        \or\chapletter={X} \or\chapletter={Y} \or\chapletter={Z}\fi}

%
\def\chapshow#1{
    \ifnum #1>0 \relax
        #1
    \else
        {\tempsetletter{\number#1}\the\chapletter}
    \fi}

%
\def\today{\number\day\space \ifcase\month\or Jan\or Feb\or
        Mar\or Apr\or May\or Jun\or Jul\or Aug\or Sep\or
        Oct\or Nov\or Dec\fi, \space\number\year}

\def\timeofday{\minutes=\time    \hours=\time
        \divide \hours by 60
        \multiply \hours by 60
        \advance \minutes by -\hours
        \divide \hours by 60
        \ifnum\the\minutes>9 \relax
     		\the\hours:\the\minutes
 	\else
  		\the\hours:0\the\minutes
	\fi}


%
%
%
%

\def\chaplabel#1{
    \ifforwardreference                             
        \write\labelfileout{                        
        \noexpand\expandafter\noexpand\def          
        \noexpand\csname CHAPLABEL#1\endcsname{\the\chapno}}
    \fi
    \global\expandafter\edef\csname CHAPLABEL#1\endcsname
    {\the\chapno}
    \ifproofmode
        \rlap{\hbox{\marginstyle #1\ }}
    \fi}

%
\def\sectnum{
    \bumpsectno
        \ifchaptersectionnumbers
            \chapfolio.
        \fi
    \the\sectno}

\def\sectlabel#1{
    \bumpsectno
    \ifforwardreference
        \immediate\write\labelfileout{
        \noexpand\expandafter\noexpand\def
        \noexpand\csname SECTLABEL#1\endcsname{\the\chapno.\the\sectno?!}}
    \fi
    \global\expandafter\edef\csname SECTLABEL#1\endcsname
    {\the\chapno.\the\sectno?!}	 			
    \ifproofmode
        \llap{\hbox{\marginstyle #1\ }}
    \fi
    \ifchaptersectionnumbers
        \chapfolio.
    \fi
    \the\sectno}

\def\sectref#1{                                  
    \ifundefined{SECTLABEL#1}                     
        ++                                        
        \ifproofmode
            \ifforwardreference
            \else
                \write16{ ***Undefined\space Section\space Reference #1*** }
            \fi
        \else
            \write16{ ***Undefined\space Section\space Reference #1*** }
        \fi
    \else
        \edef\LABxx{\getlabel{SECTLABEL#1}}
	\ifchaptersectionnumbers
            \def\LAByy{\expandafter\stripchap\LABxx}
	    \chapshow\LAByy.
	\fi
	\expandafter\stripsect\LABxx
    \fi
    \ifproofmode
        \write\allcrossfile{Section\space #1}
    \fi}

%
%
\def\eqnum{                                    
    \global\advance\eqlabelno by 1              
    \eqno(
    \ifeqchapternumbers
        \chapfolio.
    \fi
    \ifeqsectionnumbers
        \the\sectno.
    \fi
    \the\eqlabelno)}

\def\eqlabel#1{                                
    \global\advance\eqlabelno by 1              
    \ifforwardreference                     
        \immediate\write\labelfileout{\noexpand\expandafter\noexpand\def
        \noexpand\csname EQLABEL#1\endcsname
        {\the\chapno.\the\sectno?\the\eqlabelno!}}
    \fi
    \global\expandafter\edef\csname EQLABEL#1\endcsname
    {\the\chapno.\the\sectno?\the\eqlabelno!}
    \eqno(
    \ifeqchapternumbers
        \chapfolio.
    \fi
    \ifeqsectionnumbers
        \the\sectno.
    \fi
    \the\eqlabelno)
    \ifproofmode
        \rlap{\hbox{\marginstyle #1}}		
    \fi}

\def\eqalignnum{                               
    \global\advance\eqlabelno by 1              
    &(\ifeqchapternumbers
        \chapfolio.
    \fi
    \ifeqsectionnumbers
        \the\sectno.
    \fi
    \the\eqlabelno)}

\def\eqalignlabel#1{                   	
    \global\advance\eqlabelno by 1 	        
    \ifforwardreference                     
        \immediate\write\labelfileout{\noexpand\expandafter\noexpand\def
        \noexpand\csname EQLABEL#1\endcsname
        {\the\chapno.\the\sectno?\the\eqlabelno!}}
    \fi
    \global\expandafter\edef\csname EQLABEL#1\endcsname
    {\the\chapno.\the\sectno?\the\eqlabelno!}
    &(\ifeqchapternumbers
        \chapfolio.
    \fi
    \ifeqsectionnumbers
        \the\sectno.
    \fi
    \the\eqlabelno)
    \ifproofmode
        \rlap{\hbox{\marginstyle #1}}			
    \fi}

\def\dnum{                                     
    \global\advance\eqlabelno by 1              
    \llap{(	 				
    \ifeqchapternumbers
        \chapfolio.
    \fi
    \ifeqsectionnumbers
        \the\sectno.
    \fi
    \the\eqlabelno)}}

\def\dlabel#1{                                 
    \global\advance\eqlabelno by 1              
    \ifforwardreference                         
        \immediate\write\labelfileout{\noexpand\expandafter\noexpand\def
        \noexpand\csname EQLABEL#1\endcsname
        {\the\chapno.\the\sectno?\the\eqlabelno!}}
    \fi
    \global\expandafter\edef\csname EQLABEL#1\endcsname
    {\the\chapno.\the\sectno?\the\eqlabelno!}
    \llap{(
    \ifeqchapternumbers
        \chapfolio.
    \fi
    \ifeqsectionnumbers
        \the\sectno.
    \fi
    \the\eqlabelno)}
    \ifproofmode
        \rlap{\hbox{\marginstyle #1}}		
    \fi}

\def\eqref#1{\ifparenequations(\fi
    \ifundefined{EQLABEL#1}***
        \ifproofmode
            \ifforwardreference
            \else
                \write16{ ***Undefined\space Equation\space Reference #1*** }
            \fi
        \else
            \write16{ ***Undefined\space Equation\space Reference #1*** }
        \fi
    \else
        \edef\LABxx{\getlabel{EQLABEL#1}}
	\def\LAByy{\expandafter\stripsect\LABxx}
        \def\LABzz{\expandafter\stripchap\LABxx}
        \ifeqchapternumbers
            \chapshow{\LABzz}.
        \else
            \ifnum \number\LABzz=\chapno \relax
            \else
                \chapshow{\LABzz}.
            \fi
        \fi
	\ifeqsectionnumbers
	    \LAByy.
	\fi
        \expandafter\stripeq\LABxx
    \fi
    \ifparenequations)\fi
    \ifproofmode
        \write\allcrossfile{Equation\space #1}
    \fi}

%
\def\fignum{                                   
    \global\advance\figureno by 1\relax         
    \iffigurechapternumbers
        \chapfolio.
    \fi
    \the\figureno}

\def\figlabel#1{				
    \global\advance\figureno by 1\relax 	
    \ifforwardreference				
        \immediate\write\labelfileout{\noexpand\expandafter\noexpand\def
        \noexpand\csname FIGLABEL#1\endcsname
        {\the\chapno.\the\sectno?\the\figureno!}}
    \fi
    \global\expandafter\edef\csname FIGLABEL#1\endcsname
    {\the\chapno.\the\sectno?\the\figureno!}
    \iffigurechapternumbers
        \chapfolio.
    \fi
    \ifproofmode
        \llap{\hbox{\marginstyle #1\ }}\relax
    \fi
    \the\figureno}

\def\figref#1{					
    \ifundefined				
        {FIGLABEL#1}!!!!			
        \ifproofmode
            \ifforwardreference
            \else
                \write16{ ***Undefined\space Figure\space Reference #1*** }
            \fi
        \else
            \write16{ ***Undefined\space Figure\space Reference #1*** }
        \fi
    \else
        \edef\LABxx{\getlabel{FIGLABEL#1}}
        \def\LABzz{\expandafter\stripchap\LABxx}
        \iffigurechapternumbers
            \chapshow{\LABzz}.\expandafter\stripeq\LABxx
        \else \ifnum\number\LABzz=\chapno \relax
                \expandafter\stripeq\LABxx
            \else
                \chapshow{\LABzz}.\expandafter\stripeq\LABxx
            \fi
        \fi
        \ifproofmode
            \write\allcrossfile{Figure\space #1}
        \fi
    \fi}

%
%
\def\pagelabel#1{
    \ifforwardreference
        \write\labelfileout{
        \noexpand\expandafter\noexpand\def
        \noexpand\csname PGLABEL#1\noexpand\endcsname{\the\pageno}}
    \fi
    \global\expandafter\edef\csname PGLABEL#1\endcsname{\the\pageno}}

\def\pageref#1{
    \ifundefined
        {PGLABEL#1}***
        \ifproofmode
        \else
            \write16{ ***Undefined\space Page\space Reference #1*** }
        \fi
    \else
        \csname PGLABEL#1\endcsname
    \fi
    \ifproofmode
        \write\allcrossfile{Page\space #1}
    \fi}

%
\def\refnum{                                      
    \global\advance\referenceno by 1\relax         
    \the\referenceno}	                           

\def\internalreflabel#1{			
    \global\advance\referenceno by 1\relax 	
    \ifforwardreference				
        \immediate\write\labelfileout{\noexpand\expandafter\noexpand\def
        \noexpand\csname REFLABEL#1\endcsname
        {\the\chapno.\the\sectno?\the\referenceno!}}
    \fi
    \global\expandafter\edef\csname REFLABEL#1\endcsname
    {\the\chapno.\the\sectno?\the\figureno!}
    \ifproofmode
        \llap{\hbox{\marginstyle #1\hskip.5cm}}\relax
    \fi
    \the\referenceno}

\def\internalrefref#1{				
    \ifundefined				
        {REFLABEL#1}!!!!			
        \ifproofmode
            \ifforwardreference
            \else
                \write16{ ***Undefined\space Footnote\space Reference #1*** }
            \fi
        \else
            \write16{ ***Undefined\space Footnote\space Reference #1*** }
        \fi
    \else
        \edef\LABxx{\getlabel{REFLABEL#1}}
        \def\LABzz{\expandafter\stripchap\LABxx}
        \expandafter\stripeq\LABxx
        \ifproofmode
            \write\allcrossfile{Reference\space #1}
        \fi
    \fi}

%
\def\reflabel#1{\item{\internalreflabel{#1}.}}

%
\def\refref#1{\internalrefref{#1}}

\def\eq{\ifhmode Eq.~\else Equation~\fi}		
\def\eqs{\ifhmode Eqs.~\else Equations~\fi}

%
%
%
%

%
\def\getlabel#1{\csname#1\endcsname}
\def\ifundefined#1{\expandafter\ifx\csname#1\endcsname\relax}
\def\stripchap#1.#2?#3!{#1}			
\def\stripsect#1.#2?#3!{#2}			%
\def\stripeq#1.#2?#3!{#3}			
}  

\overfullrule = 0pt
\magnification = 1200
\baselineskip 21pt
\forwardreferencetrue
\initialeqmacro

\def\ch{\mathop{\rm ch}\nolimits}


\line{\hfill LPTHE-PAR 93-09, UMTG-171}

\vskip 0.2 in

\centerline{\bf Thermodynamics of Integrable Chains with Alternating Spins}
\bigskip

\medskip

\centerline{H. J. de Vega\footnote{$\dagger$}{Laboratoire de Physique
Th\'eorique et Hautes Energies, Tour 16 - 1er. \'etage, Universit\'e
Paris VI, 4, place Jussieu, 75252 Paris Cedex 05, FRANCE},
Luca Mezincescu${}^*$, and
Rafael I. Nepomechie\footnote*{Department of Physics, University of Miami,
Coral Gables, FL 33124, USA}}

\vskip 0.2 in

\bigskip

\centerline{\bf Abstract}

\vskip 0.2 in

We consider a two-parameter $( \bar c \,, \tilde c )$ family of quantum
integrable Hamiltonians for a chain of alternating spins of spin $s=1/2$ and
$s=1$. We determine the thermodynamics for low-temperature $T$ and small
external magnetic field $H$, with $T << H$.
In the antiferromagnetic $( \bar c > 0\,, \tilde c > 0)$ case, the model has
two gapless excitations. In particular,
for $\bar c = \tilde c$, the model is conformally invariant and has central
charge $c_{vir} = 2$. When one of these parameters is zero, the Bethe
Ansatz equations admit an infinite number of solutions with lowest energy.

\vfill\eject

The one-dimensional Heisenberg model, like the hydrogen atom, has served
as the guiding example for a very large body of both experimental and
theoretical work. Progress has recently been made on closely related models,
consisting of chains with {\it alternating} spins, such as spin 1/2 and spin 1.
On the experimental side, materials (e.g.${}^{\refref{experimental}}$,
[MnCp${}_2^*$] [TCNE]) have been synthesized which, at
temperatures above a certain transition temperature $T_c$, behave as
one-dimensional ferromagnets of alternating spins.
On the theoretical side, quantum integrable models of chains with alternating
spins have recently been constructed${}^{\refref{devega/woynarovich}}$. In this
Letter we investigate the thermodynamics of a two-parameter family of such
integrable models. Depending on the values of the parameters, we find both
antiferromagnetic and ferromagnetic behavior. When one of these parameters is
zero, the Bethe Ansatz equations admit an infinite number of solutions with
lowest energy.

We consider a system of $N$ spins
${1\over 2}\vec \sigma_2 \,, {1\over 2}\vec \sigma_4 \,, \cdots \,,
{1\over 2}\vec \sigma_{2N}$
of spin 1/2 and $N$ spins $\vec s_1 \,, \vec s_3 \,,$  \break
$\cdots \,, \vec s_{2N-1}$  of spin 1 in an external magnetic field
$H (\ge 0)$ with the Hamiltonian ${\cal H}$ given by
$$ {\cal H} = \bar c  \bar {\cal H} + \tilde c \tilde {\cal H}
- H S^z \,, \eqlabel{hamiltonian} $$
where
$S^z =\sum_{n=1}^N {1\over 2} \sigma^z_{2n} + \sum_{n=1}^N s^z_{2n-1}$,
$$\eqalignno{
  \bar {\cal H} &= -{1 \over 9} \sum_{n=1}^{N} \left( 2 \vec \sigma_{2n} \cdot
\vec s_{2n+1} + 1 \right) \left( 2 \vec \sigma_{2n+2} \cdot
\vec s_{2n+1} + 3 \right) \,, \eqalignnum \cr
\tilde {\cal H} &=  -{1 \over 9} \sum_{n=1}^{N} \left( 2 \vec \sigma_{2n} \cdot
\vec s_{2n-1} + 1 \right) \left[ \left( 1 +  \vec s_{2n-1} \cdot
\vec s_{2n+1}\right) \left( 2 \vec \sigma_{2n} \cdot
\vec s_{2n+1} + 1 \right) + 2 \right] \,, \eqalignnum \cr} $$
and $\bar c$ and $\tilde c$ are real constant parameters. (In this paper,
bars and tildes are interchanged with respect to Ref.
\refref{devega/woynarovich}.) Note that the Hamiltonian contains both nearest
and next-to-nearest neighbor interactions. We assume periodic boundary
conditions: $\sigma_{2n} \equiv \sigma_{2n + 2N}$ and $s_{2n+1}
\equiv s_{2n +1 + 2N}$.

The corresponding energy eigenvalues are given
by${}^{\refref{devega/woynarovich}}$
$$ E = \bar c  \bar E + \tilde c \tilde E  - H \left( {3\over 2}N - M  \right)
\,, \eqlabel{eigenvalues} $$
where
$$
\bar E = {5\over 3}N - i  \sum_{j=1}^M {d\over d\lambda_j}
\ln \left( \lambda_j + {i\over 2} \over \lambda_j - {i\over 2} \right) \,,
\qquad
\tilde E = - {1\over 2}N - i  \sum_{j=1}^M {d\over d\lambda_j}
\ln \left( \lambda_j + i \over \lambda_j - i \right) \,,
\eqlabel{EE}  $$
where the variables $\lambda_j$ satisfy the Bethe Ansatz (BA) equations
$$
\left( {\lambda_j + {i\over 2} \over \lambda_j - {i\over 2}}
       {\lambda_j + i \over \lambda_j - i} \right)^N
= - \prod_{k=1}^M {\lambda_j - \lambda_k + i \over
\lambda_j - \lambda_k - i } \,, \qquad j = 1, \cdots , M \,. \eqlabel{BA} $$
We consider here a strictly alternating arrangement of spins, with
spins 1/2 at even sites and spins 1 at odd sites. For any other ordering
of the spins, one can construct a corresponding Hamiltonian which has the
same energy eigenvalues and BA equations.

This system of equations admits the same string solutions that are
found for the Heisenberg model. In the thermodynamic limit, the model
is characterized by particle densities $\rho_n (\lambda)$ and hole densities
$\tilde \rho_n (\lambda)$. Following the standard procedure (see, e.g.,
Refs. \refref{tsvelick/wiegmann}, \refref{xxx/spin/s}),
we find that these densities obey the constraints
$$\tilde \rho_n +  \sum_{m=1}^\infty A_{nm} * \rho_{m} = a_n
+ \sum_{l=1}^{min(n,2)} a_{n+3 -2l} \,, \eqlabel{constraint}$$
where
$$a_n(\lambda) = {1\over 2\pi} {n\over \lambda^2 + {n^2\over 4}} \,, \eqnum $$
$$A_{nm}(\lambda) = \delta_{nm}\delta(\lambda)
+ \left(1 - \delta_{nm} \right) a_{|n - m|}(\lambda) + a_{n + m}(\lambda)
+ 2\sum_{l=1}^{min(n,m)-1} a_{|n - m| + 2l}(\lambda) \,, \eqlabel{Anm} $$
and $*$ denotes the convolution
$ \left( f * g \right) (\lambda) = \int_{-\infty}^\infty
d\lambda'\ f(\lambda - \lambda') g(\lambda') \,.$

The thermodynamic Bethe Ansatz (TBA) equations describing the equilibrium
system at temperature $T$ are
$$T \ln \left( 1 + e^{\epsilon_n/T} \right)
= \sum_{m=1}^\infty  A_{nm} * T \ln \left( 1 + e^{-\epsilon_m/T} \right)
-2\pi \left( \bar c a_n + \tilde c \sum_{l=1}^{min(n,2)} a_{n+3-2l} \right)
 + n H \,, \eqlabel{TBA} $$
where
$$\epsilon_n (\lambda) = T \ln \left( \tilde\rho_n (\lambda)/
\rho_n (\lambda)\right) \,. \eqlabel{epsilon}  $$
(The particle and hole densities are understood to be those at equilibrium.)
The equilibrium free energy is given by
$$F = 2N e_0
- N T \int_{-\infty}^\infty  d\lambda\ s(\lambda)
\ln \left[ \left( 1 + e^{\epsilon_1(\lambda)/T} \right)
           \left( 1 + e^{\epsilon_2(\lambda)/T} \right) \right]
\,, \eqlabel{free} $$
where $s(\lambda) = 1/ \left( 2 \ch \pi \lambda \right) \,,$
and
$ 2 e_0 = \bar c \left[ {5\over 3}  - \left( 2\ln 2 + 4 -\pi \right) \right]
       + \tilde c \left[ - {1\over 2} - \left( 6 - \pi \right) \right]
 \,.$

In the high-temperature limit with zero field, these equations give the
expected value for the entropy, namely $S = N \ln 6$. We now
consider small values of $T$ and $H$, with $T << H$. We define
$\varepsilon_n(\lambda) = \lim_{T \rightarrow 0}\ \epsilon_n(\lambda) \,.$
Keeping in mind that $\epsilon_n > 0$ for $H > 0$ and $n \ge 3$, we obtain from
the TBA equations the following system of linear integral equations for
$\varepsilon_1$ and $\varepsilon_2$:
$$\eqalignno{
\varepsilon_1 &= -2\pi \bar c s + s * \varepsilon^+_2  \,, \cr
\varepsilon_2 &= {1\over 2}H -2\pi \tilde c s + h * \varepsilon^+_2
+ s * \varepsilon^+_1 \,, \eqalignlabel{varepsilon} \cr} $$
where $h = s * a_1 \,,$
and the superscript $+$ on a function denotes the positive part
of that function; i.e.,
$\varepsilon^+ \equiv {1\over 2} \left( \varepsilon + |\varepsilon| \right)
\,.$

The qualitative behavior of the solutions depends on the sign of the
parameters $\bar c$ and $\tilde c$, and the various cases must be
studied individually. Let us first consider the
antiferromagnetic case $\bar c > 0$, $\tilde c > 0$. In this case, for $H = 0$
the solutions $\varepsilon_1(\lambda)$ and $\varepsilon_2(\lambda)$ are
readily found, and are seen to be negative for all $\lambda$. Hence, for the
ground state, $\rho_1(\lambda) = \rho_2(\lambda) = s(\lambda) \,,$
and all other particle and hole densities are equal to
zero${}^{\refref{note1}}$.
Thus, the ground state is a ``sea'' of strings of length 1 and a
``sea'' of strings of length 2, in agreement with the alternative analysis of
Ref. \refref{devega/woynarovich}. This corresponds to the antiferromagnetic
ground states of integrable chains${}^{\refref{xxx/spin/s}}$ of
spin $s=1/2$ and $s=1$.

For $H$ small but nonzero, we can solve for $\varepsilon_1$ and $\varepsilon_2$
by generating from \eqref{varepsilon} a system of Wiener-Hopf equations.
To this end, we define $\alpha_n$ to be the zeros of
$\varepsilon_n(\lambda)$, i.e.,
$ \varepsilon_n(\alpha_n) =  0 \,,$  for $n = 1\,, 2 \,.$
We assume
$$\alpha_n = -{1\over \pi} \left[
\ln H  + \ln \kappa_n + O \left( {1\over \ln H} \right) \right]
\,, \eqlabel{kappa} $$
where the constants $\kappa_n$ (which are independent of $H$) have still
to be determined. Introducing the functions
$$S_n(\lambda) = \left\{ \matrix{e^{\pi \alpha_n}
\kappa_n \varepsilon_n (\lambda + \alpha_n)  & \lambda > 0 \cr
0 & \lambda < 0 \cr} \right. \,, \eqlabel{funcS} $$
we obtain (for $H \rightarrow 0$) the following system of Wiener-Hopf equations
$$\eqalignno{
S_1(\lambda) &= -2\pi \bar c \kappa_1 e^{-\pi\lambda}
+ \int_0^\infty d\lambda'\
s(\lambda - \lambda' + \alpha_1 - \alpha_2) S_2(\lambda') \,, \cr
S_2(\lambda) &= {1\over 2} -2\pi \tilde c \kappa_2 e^{-\pi\lambda}
+ \int_0^\infty d\lambda'\ \left[
s(\lambda - \lambda' + \alpha_2 - \alpha_1) S_1(\lambda') +
h(\lambda - \lambda') S_2(\lambda') \right] \,, \cr
&\qquad\qquad\qquad\qquad\qquad\qquad\qquad\qquad\qquad\qquad
\qquad\qquad\qquad\qquad \lambda \ge 0 \,.
\eqalignlabel{WHS} \cr} $$

In order to find the leading-order temperature dependence of the free energy,
we must compute the leading correction to the solutions
$\epsilon_n = \varepsilon_n$ of the linearized equations \eqref{varepsilon}.
Hence, we set
$\epsilon_n (\lambda) = \varepsilon_n(\lambda) + \eta_n (\lambda)$
in the TBA equations and expand to leading order in $\eta_n$, as is explained
in Refs. \refref{johnson/mccoy} and \refref{kingston}.
In terms of the functions
$$T_n(\lambda) = \left\{ \matrix{ {6 e^{-\pi \alpha_n}
\over \pi^2 T^2 \kappa_n}
\eta_n (\lambda + \alpha_n)  & \lambda > 0 \cr
0 & \lambda < 0 \cr} \right. \,, \eqlabel{funcT} $$
we obtain (for $H \rightarrow 0$) a second system of Wiener-Hopf equations:
$$\eqalignno{
T_1(\lambda) &=
{s(\lambda + \alpha_1 - \alpha_2)\over S_2'(0)}
+ \int_0^\infty d\lambda'\
s(\lambda - \lambda' + \alpha_1 - \alpha_2) T_2(\lambda') \,, \cr
T_2(\lambda) &= {h(\lambda)\over S_2'(0)} +
{s(\lambda + \alpha_2 - \alpha_1)\over S_1'(0)}
+ \int_0^\infty d\lambda'\ \left[
s(\lambda - \lambda' + \alpha_2 - \alpha_1) T_1(\lambda') +
h(\lambda - \lambda') T_2(\lambda') \right] \,, \cr
&\qquad\qquad\qquad\qquad\qquad\qquad\qquad\qquad\qquad\qquad
\qquad\qquad\qquad\qquad \lambda \ge 0 \,,
\eqalignlabel{WHT} \cr} $$
where $ S_n'(0) = d S_n/ d\lambda \Big\vert_{\lambda = 0^+}$.

Both systems \eqref{WHS} and \eqref{WHT} involve the same
$2 \times 2$ matrix kernel, which in Fourier space is given by
$$\hat K (\omega) = \left(
\matrix{ 0 &  e^{-i \omega
\left( \alpha_1 -\alpha_2 \right)} \hat s(\omega)\cr
e^{i\omega \left( \alpha_1 - \alpha_2 \right)}
\hat s(\omega)  & \hat h(\omega) \cr} \right) \,.
\eqlabel{kernel} $$
(Our convention for Fourier transforms is that
$\hat f(\omega) = \int_{-\infty}^\infty d\lambda\ e^{i \lambda \omega}
f(\lambda)$.)
Results from Ref. \refref{gohberg/krein} imply that the following factorization
exists
$$ \left( 1 - \hat K(\omega) \right)^{-1} = G_+ (\omega)\ G_- (\omega)
\,, \qquad\qquad -\infty < \omega < \infty \,,
\eqlabel{factorization} $$
where $G_+ (\omega)$ and $G_+^{-1} (\omega)$ are analytic in the upper-half
complex $\omega$ plane with $G_+ (+\infty) = 1$, and (for $\omega$ in the
lower-half plane) $G_-(\omega) = G_+(-\omega)^T \,.$

Using standard Wiener-Hopf methods, we  conclude that the solutions of
Eqs. \eqref{WHS} and \eqref{WHT} are given (in matrix notation) by
$$\hat S(\omega) = {i\over 2}
\left( {1\over \omega + i0} - {1\over \omega + i\pi} \right)
G_+(\omega) G_-(0) \left( \matrix{ 0 \cr
                                   1 \cr} \right)
 \,, \eqlabel{Ssolution} $$
and
$$\hat T (\omega) = \left( G_+(\omega) - 1 \right)
\left( \matrix{ 1/S_1'(0) \cr
                1/S_2'(0) \cr} \right) \,, \eqlabel{Tsolution} $$
where
$$ S'(0) = - \lim_{|\omega| \rightarrow \infty} \omega^2 \hat S(\omega)
= {\pi \over 2}
G_-(0) \left( \matrix{ 0 \cr
                       1 \cr} \right)
\,. \eqlabel{derivative} $$
Moreover, the parameters $\kappa_n$ introduced in \eq\eqref{kappa} are given by
$$\kappa = \left( \matrix{ \kappa_1 \cr
                           \kappa_2 \cr} \right)
= {1\over 4\pi} \left( \matrix{ 1/\bar c & 0 \cr
                                  0      & 1/\tilde c \cr} \right)
G_-(-i\pi )^{-1} G_-(0) \left( \matrix{ 0 \cr
                                        1 \cr} \right) \,.
\eqnum $$

In order to calculate the free energy per site $f= F/2N$, we substitute
$\epsilon_n = \varepsilon_n + \eta_n$ into the expression \eqref{free}
for the free energy, as is explained in Refs. \refref{johnson/mccoy}
and \refref{kingston}. We obtain
$$f = e_0 - H^2 A  - {\pi^2 T^2 \over 6} B \,, \eqnum $$
with
$$ A = \kappa^T\ \hat S(i\pi) \,, \qquad
   B = \kappa^T\ \left[ \hat T(i\pi)
+ \left( \matrix{ {1/ S_1'(0)} \cr
                  {1/ S_2'(0)} \cr}  \right) \right]  \,.
\eqnum $$

For $\bar c \ne \tilde c$, the quantities $A$ and $B$ cannot be explicitly
evaluated without having explicit expressions for the factors
$G_+(\omega)$ and $G_-(\omega)$, which we have not yet found.

For the special case $\bar c = \tilde c \equiv c$, the quantities $A$ and $B$
can be readily evaluated, and we conclude (for $T << H$ )
$$f = e_0 - {1\over 4\pi^2 c} H^2 - {1\over 6 c}T^2 \,. \eqlabel{final} $$
It follows that the magnetic susceptibility and specific heat, to lowest
order, are given by
$$
\chi = - {\partial^2 f\over \partial H^2}
\Big\vert_T = {1\over 2\pi^2 c} \,, \qquad
C_H = -T {\partial^2 f\over \partial T^2}
\Big\vert_H = {1\over 3 c}T \,,   \eqnum $$
respectively.

This model has${}^{\refref{devega/woynarovich}, \refref{note}}$ two gapless
excitations, with corresponding speeds of sound $\bar v = 2\pi \bar c$ and
$\tilde v = 2\pi \tilde c$. Evidently, the case $\bar c = \tilde c$ is
the unique case for which the two speeds of sound coincide, and
the model is conformally invariant. For a critical chain, the low-temperature
free energy per site is given by${}^{\refref{cardy}, \refref{affleck1}}$
$$f = e_0 - {\pi c_{vir}\over 6 v_s} T^2 + \cdots \,, \eqnum $$
where $c_{vir}$ is the central charge of the Virasoro algebra and $v_s$ is
the speed of sound. Therefore, from \eqref{final} we see that $c_{vir} = 2$.
Presumably it is no coincidence that precisely for the conformally invariant
case, explicit expressions for $G_+(\omega)$ and $G_-(\omega)$ are not
needed to evaluate the free energy.

We now consider the case $\bar c = 0$, $\tilde c > 0$. For this case, there
is a one-parameter $(\alpha)$ family of lowest-energy states. Indeed,
consider the following one-parameter family of densities:
$$
\rho_1 = \alpha s \,, \qquad \tilde \rho_1 = (1 - \alpha) s \,, \qquad
\rho_2 = s + (1 - \alpha) s*s \,,  \qquad\qquad
0 \le \alpha \le 1 \,, \eqnum $$
and all other particle and hole densities are equal to zero. These densities
obey the constraints \eqref{constraint}, and give (independently of
the value of $\alpha$) the same lowest value for the energy. Moreover, to
leading order in $N$, the spin is $S^z = 0$, and the entropy is
$S = -{N\over 2}\left[ \alpha \ln \alpha + (1 - \alpha )\ln (1 - \alpha)
\right]$. In particular, for $\alpha \ne 0 \,, 1$,  the entropy is nonzero
and is proportional to $N$, implying an infinite degeneracy of states.
This degeneracy is consistent with the fact that, above the $\alpha =1$
vacuum, there are excitations (namely, holes in the sea of real roots) which
have {\bf zero energy} and non-zero momentum${}^{\refref{devega/woynarovich}}$.
We are not aware of any other model with such properties${}^{\refref{note2}}$.

We speculate that the system can be brought to these various states by first
preparing the system at finite $T$ and $H$, and then approaching the origin
$(T=0 \,, H=0)$ of the $(T \,, H)$-plane from appropriate directions.
The ground state of the system is the state which is reached by approaching
the origin along the line $H=0$. Unfortunately, we cannot determine
the particular value of $\alpha$ corresponding to this state,
since this would entail computing
$\lim_{T \rightarrow 0}\ \lim_{H \rightarrow 0} \epsilon_1/T$, while
we know how to calculate only for $T << H$ ${}^{\refref{note3}}$.
By approaching the origin of the $(T \,, H)$-plane along the line $T=0$,
the state with $\alpha = 0$ is reached.

We have calculated the free energy for small values of $T$ and $H$,
with $T << H$. The calculation is similar to the one above,
except that now one must take into account that $\varepsilon_1 (\lambda)$
does not have a zero. We find that, to leading order, the free energy per site
is given by
$$f = e_0 +  {1\over 4\pi^3 \tilde c} H^2 \ln H +
{1\over 12 \pi \tilde c} T^2 \ln H \,. \eqnum $$
Contrary to appearance, this result does {\it not} imply that
$f$ diverges for $H \rightarrow 0$, since in the region where the calculation
is valid ($T << H$), the last term is finite. This result is nevertheless
unusual, since it implies that $C_H/T$ is a function of $H$.

Similar results are obtained for the case $\tilde c = 0$, $\bar c > 0$.
For the cases $\bar c < 0$, $\tilde c > 0$ and $\bar c > 0$, $\tilde c < 0$ ,
the model either has a ferromagnetic ground state and a finite gap, or
it has no gap, depending on the precise values of $\bar c \,, \tilde c \,,$
and $H$.
For the cases $\bar c \,, \tilde c \le 0 \,,$ the model is ferromagnetic.

\bigskip

We thank L. Susskind for a very helpful discussion and F. Zuo for bringing
Ref. \refref{experimental} to our attention. One of us (H J de V) thanks the
Department of Physics of the University of Miami for the warm hospitality
extended to him. This work was supported in part by the National Science
Foundation under Grant No. PHY-92 09978.

\bigskip

Note added: After this work was completed, we received a preprint
(Ref. \refref{martins}) which discusses chains of alternating spin 1/2 and
spin $s$. However, that paper considers neither the effect of an external
magnetic field, nor the behavior of the models away from the conformally
invariant point (i.e., for $\bar c \ne \tilde c$).

\bigskip

\noindent
{\bf References}

\vskip 0.2truein

\reflabel{experimental}
G.T. Yee, J.M. Manriquez, D.A. Dixon, R.S. McLean, D.M. Groski, R.B. Flippen,
K.S. Narayan, A.J. Epstein, and J.S. Miller, Adv. Mat. {\it 3}, 309 (1991).

\reflabel{devega/woynarovich}
H.J. de Vega and F. Woynarovich, J. Phys. {\it A25}, 4499 (1992).

\reflabel{tsvelick/wiegmann}
A.M. Tsvelick and P.B. Wiegmann, Adv. in Phys. {\it 32}, 453 (1983).

\reflabel{xxx/spin/s}
H.M. Babujian, Nucl. Phys. {\it B215}, 317 (1983);
L.A. Takhtajan, Phys. Lett. {\it 87A}, 479 (1982).

\reflabel{note1}
By definition, the set of ground state densities should be determined through
\eq\eqref{epsilon} by computing
$\lim_{T \rightarrow 0}\ \lim_{H \rightarrow 0} \epsilon_n/T$,
and then imposing the constraints \eqref{constraint}. Here, we compute
instead $\lim_{H \rightarrow 0}\ \lim_{T \rightarrow 0} \epsilon_n/T$, and
we assume that the two limits commute. If the limits did not commute,
there would be another set of densities which would also satisfy the
constraints \eqref{constraint} and give the same value for the
energy. We believe that for the case $\bar c > 0$, $\tilde c > 0$
such an additional set of densities does not exist.

\reflabel{johnson/mccoy}
J.D. Johnson and B.M. McCoy, Phys. Rev. {\it A6}, 1613 (1972).

\reflabel{kingston}
L. Mezincescu and R.I. Nepomechie, in {\it Quantum Groups, Integrable Models
and Statistical Systems}, ed. by J. Le Tourneux and L. Vinet (World
Scientific), in press.

\reflabel{gohberg/krein}
I.C. Gohberg and M.G. Krein, in {\it American Mathematical Society
Translations, Series 2, Vol 14} (American Mathematical Society, 1960) 217.

\reflabel{note}
We define the momentum here as one-half the log of the two-site shift operator.
This yields the stated values for the speeds of sound, and a system of length
$2N$ (see Eq. (25)). In Ref. 2, the log of the two-site shift operator is used,
which leads to half these values for the speeds and length. We thank M. Martins
for a discussion on this point.

\reflabel{cardy}
H.W.J. Bl\"ote, J.L. Cardy and M.P. Nightingale,
Phys. Rev. Lett. {\it 56}, 742 (1986).

\reflabel{affleck1}
I. Affleck, Phys. Rev. Lett. {\it 56}, 746 (1986).

\reflabel{note2}
Implicit in this discussion is the conventional assumption that a set of
densities which satisfies the constraints \eqref{constraint} and has lowest
energy corresponds to an eigenstate of the Hamiltonian.

\reflabel{note3}
In contrast to the case $\bar c > 0$, $\tilde c > 0$, here we expect that
$\lim_{T \rightarrow 0}\ \lim_{H \rightarrow 0} \epsilon_1/T \ne
 \lim_{H \rightarrow 0}\ \lim_{T \rightarrow 0} \epsilon_1/T \,. $

\reflabel{martins}
S.R. Aladim and M.J. Martins, ``Critical behavior of integrable mixed
spin chains'' UFSCAR-93-03.

\end